\documentstyle[12pt]{article}
\include{./psfig}
\def\eqn{\begin{equation}}
\def\enn#1{\label{#1} \end{equation}}
\def\eq{\[}
\def\en{\]}
\def\r#1{eq. (\ref{#1})}

\def\f#1{Fig. {\ref{#1}}}
\def\t#1{Table \ref{#1}}
\def\p#1{\ref{#1}}

\def\Mu{{\cal{M}}}

\begin{document}

\pagestyle{plain}
\title{Coarse Projective kMC Integration:
Forward/Reverse Initial and Boundary Value Problems}

\author{R. Rico-Mart\'{\i}nez$^{1,2}$, C. W. Gear$^{1,3}$ and
Ioannis G. Kevrekidis$^{1,4}$\\
$^1$Princeton  University, Dept. of Chemical Engineering\\
$^2$Instituto Tecnol\'ogico de Celaya, Dpto. de Ingenier\'{\i}a Qu\'{\i}mica\\
$^3$NEC Laboratories (retired)\\
$^4$Princeton  University, Applied and
Comput. Mathematics/Mathematics}

\maketitle

\begin{abstract}

In ``equation-free" multiscale computation a dynamic model is given
at a fine, microscopic
level; yet we believe that its coarse-grained, macroscopic dynamics
can be described by closed equations involving only coarse variables.
These variables are typically various low-order moments of the
distributions evolved through the microscopic model.
We consider the problem of integrating these unavailable equations
by acting directly on kinetic Monte Carlo microscopic
simulators, thus circumventing their derivation in closed form.
In particular, we use projective multi-step integration to
solve the coarse initial value problem forward in time
as well as backward in time (under certain conditions).
Macroscopic trajectories are thus {\it traced back} to unstable,
source-type, and even sometimes saddle-like stationary points,
even though the microscopic simulator only evolves forward in
time.
We also demonstrate the use of such projective integrators in
a shooting boundary value problem formulation
for the computation of ``coarse limit cycles" of the macroscopic
behavior, and the approximation of their stability through
estimates of the leading ``coarse Floquet multipliers".

\end{abstract}

{\bf Keywords:} Projective Integration, Kinetic Monte Carlo.

{\bf Mathematical Reviews Index Classification:} 65C05. Numerical Simulation,
Monte Carlo Methods.

\section{Introduction}

In previous work (introduced in \cite{PNAS} and developed in a sequence
of publications
\cite{alexi,alexei2,Siettos,Siettos2,cwgygk,gkt,Runborg,prl,smms})
whose underlying principles are discussed
in detail in \cite{Manifesto}, we have used {\em coarse time-stepping}
as a tool
for the computer-assisted analysis (under appropriate
conditions) of macroscopic process evolution,
even when the model of the process is only known at a fine (atomistic,
stochastic, microscopic) level.
When coarse-grained, closed descriptions exist, but are not available in
closed form, our so-called ``equation free" approach may provide a framework
for bridging microscopic modeling and traditional continuum numerical analysis.
The quantities necessary for ``systems level" numerical tasks are
{\it estimated} using short bursts of appropriately initialized
microscopic simulations, and tools from systems theory (identification,
filtering, variance reduction etc.).
This ``closure on demand" procedure, coupled with matrix-free techniques
of modern iterative large scale linear algebra, enables microscopic
simulators to directly perform tasks like accelerated integration,
fixed point computation, stability/parametric and bifurcation analysis,
controller design and optimization directly, without
ever obtaining the coarse level equations in closed form.
The extraction of macroscopic dynamics from microscopic simulators is a
subject of intense current research interest in disciplines ranging
from mathematics to materials science, and from computational biology
to weather prediction.
An extensive discussion with references can be found in \cite{Stuart}; we
mention here, in particular, the early work of Chorin and coworkers on optimal
predictors (\cite{Chorin1,Chorin2}.

A coarse timestepper is based on a pair of
transformations between the microscopic (fine) and macroscopic (coarse)
descriptions: (a) {\em lifting}, $\mu$, which takes a macroscopic
initial state into consistent (usually higher-dimensional) microscopic
descriptions; and (b) {\em restriction}, $\Mu$, which goes in the other
direction, giving macroscopic observations of detailed microscopic states.
A coarse time step effectively computes the change
in the expected, macroscopic description over a time horizon $\Delta T =
n\delta t$ where $\delta t$ is the time step of the microscopic
model.
One coarse time step starting from the macroscopic initial condition
$Y(T)$ consists of
\begin{enumerate}
\item Lifting to one (or possibly an ensemble of) consistent
initial conditions at the microscopic level: $y(T) = \mu
Y(T)$
\item Integrating forward (evolving) at the microscopic
level for $n$ time steps to $y(T+n\delta t)$.
\item Restricting the final answer to the coarse variables $Y(T+\Delta T)
= \Mu y(T+n\delta t)$.
\end{enumerate}

Applied directly to long time simulation, the coarse time-stepper
would do nothing to reduce the cost of computing at the
microscopic level.
Neither would there be any point to performing
lifting operations after the initial one, since detailed microscopic states
are available from the end of the previous coarse time step.
It is when coarse time-stepping is used in conjunction with other
techniques that it provides computational and analytical benefits.
These techniques are based on the observation that estimates of certain
additional quantities (e.g. of the time-derivative of the coarse
evolution, or even Fr\'echet derivatives, i.e. the action of the linearization
of the coarse evolution) can be relatively systematically obtained.
The time-derivative of the coarse solution can be estimated through
the coarse time-stepper using (in the simplest of several
approaches) the chord of the solution over a time interval $N \Delta T$
for integer $N$.
Thus, stationary state problems can be approximately solved by finding
a state, $Y^*$, such that the chord slope is zero.
More importantly for the purpose of this paper, the
chords can be used as inputs to an {\em outer integrator} of the
macroscopic variables in a process we called {\em coarse projective
integration} \cite{cwgygk}.
In the same spirit, a coarse limit cycle can be found by solving
a two-point boundary value problem at the macroscopic level
using the same coarse projective integrators in a shooting
formulation - i.e., by converging on the fixed point of a
coarse Poincar\'e map.

In computing the chord slope we do not, usually, take just one
coarse step, but perform several preliminary ones after the
initial lifting operation, and then compute the chord slope of the
final one.
This is done to allow the initial lifted values to
``heal" - that is, to allow for higher-order moments
of the microscopically evolving distributions to get
``slaved to" (i.e. become functionals of) the lower-order
``master" moments used to parametrize the coarse description.
The basic assumption is that an attracting ``slow manifold"
underpins the coarse dynamics; this manifold is parametrized
by a set of ``coarse variables" (typically the first few moments
of the microscopically evolving distributions).
In principle, the expected values of the remaining moments can be
plotted as an unspecified function of the coarse variables;
if detailed simulations are initialized away from this manifold,
they quickly evolve towards it and then approximately ``on it".
A simple but important observation is that in general,
if this ``slaving" does not occur quickly (compared to the
observation time of the simulator/experimenter), the system
cannot be deterministically described in terms of the current
set of coarse variables; this means that no deterministic
macroscopic equations exist and close at this level of description.
This picture, and its association with the ideas of approximate inertial
manifolds \cite{Foias,Temam} is discussed in more detail in
\cite{Manifesto,gkt,alexi,Hummer}.

Since (in the simplest implementation) the values of the macroscopic
variables are only needed for
the chord computation, the restriction operation only need be
performed at the two points needed to compute the chord.
Thus the
chord calculation process to find the slope, $S$, of a chord of the solution
through the points $Y(T+n_1\delta t)$ and $Y(T+n_2\delta t)$ starting
from $Y(T)$ is:
\begin{enumerate}
\item Lift from $Y(T)$ to (possibly several) consistent $y(T)$
\item Perform $n_1$ microscopic simulation steps to get $y(T+n_1\delta
t)$.  This is the {\em settling} or {\em healing time}
that allows the initially
lifted distribution to ``settle" to a realistic distribution
before estimating the slope.
Alternatively, this is the time for the simulations to approach the
coarse slow manifold (i.e. for the higher order moments to become functionals
of the lower order, governing moments).
\item Restrict to get $Y(T+n_1\delta t)$.
\item Perform $n_2 - n_1$ more microscopic simulation steps to get
$y(T+n_2\delta t)$.
\item Restrict to get $Y(T+n_2\delta t)$.
\item Compute $S = (Y(T+n_2\delta t)-Y(T+n_1\delta
t))/((n_2-n_1)\delta t)$.
\end{enumerate}

If the microscopic model is stochastic, the chord slope of the
coarse time-stepper will be noisy.
The variance of this
noise can be reduced in several ways: the size (number of
particles or other components) of the microscopic model can be
increased, multiple copies of the microscopic model can be run
(in parallel!) and the results averaged, and/or - the method we will
use here - the chord can be computed by fitting a straight line to
the output of a number of (coarse) time steps.
Other approaches to variance reduction are discussed in
\cite{oettinger1,oettinger2}.
In our case the chord calculation process is
\begin{enumerate}
\item Lift from $Y(T)$ to $y(T)$
\item Perform $n_1$ microscopic simulation steps to get $y(T+n_1\delta
t)$.
\item Restrict to get $Y(T+n_1\delta t)$.
\item Repeat the next two steps for $q = 1, 2, \cdots, m$
\item Perform $n_3$ more microscopic simulation steps to get
$y(T+(n_1+qn_3)\delta t)$.
\item Restrict to get $Y(T+(n_1+qn_3)\delta t)$.
\item Compute $S$ as the slope of the least-squares linear fit to
$Y(T+(n_1+qn_3)\delta t)$ for $q = 1, 2, \cdots, m$.
\end{enumerate}
The process is illustrated in \f{fig1}. In Sections 2 and 3 of
this paper we will use this form of chord computation process to
perform high-order multistep integration and limit cycles,
respectively.

The model problem considered here was chosen (for validation
purposes) to be one for which we do know the coarse-grained
macroscopic equations - so that comparisons can be made between
a coarse stochastic integration based on the microscopic model and the
``true" deterministic macroscopic equations for the expected dynamics.
We chose one of the examples used in \cite{alexi}; it is a kinetic Monte Carlo
(kMC) realization (using the stochastic simulation algorithm
of Gillespie \cite{Gillespie_1,Gillespie_2})
of a simple surface reaction model for which the mean
field evolution equation for the surface coverages ($\theta_i$) of the
participating species are known \cite{alexi}.

\begin{eqnarray}
{{d\theta_A}\over{dt}} & = &  \alpha \theta_*-\gamma
\theta_A-4k_r\theta_A\theta_B \nonumber \\
{{d\theta_B}\over{dt}} & = &  2 \beta  \theta_*^2
-4k_r\theta_A\theta_B \nonumber \\
{{d\theta_C}\over{dt}} & = &  \mu \theta_*-\eta\theta_C  \label{alexieq}
\end{eqnarray}

\noindent
This is a simplified model of the oxidation of CO ($A$) by dissociatively
adsorbing oxygen ($B$) on a Pt catalyst surface in the presence of an
additional inert species ($C$).
Here $\theta_*=1-\theta_A-\theta_B-\theta_C$.
For our computations the
values of the parameters are set as follows: $\mu=0.36$, $\eta=0.016$,
$\alpha=1.6$, $\gamma=0.04$, $k_r=1.0$; $\beta$ was set as cited
in the text.

The kMC simulations were performed as described in \cite{alexi} using
$N=(1000)^2$ adsorption sites and computing the average over several
realizations, typically 100.  Comparisons were made with numerical
integration of the deterministic model \r{alexieq} using
the implicit solver ODESSA \cite{odessa}.

In the next Section we will describe the multi-step coarse projective
integration method we use and some preliminary tests that were made to
decide on an appropriate order for the projective method.
In Section 3 we will discuss the computation of coarse limit cycles
through shooting with a coarse projective integrator.
In Section 4 we will show how to perform reverse coarse projective
integration through forward kMC simulation, and show that the
reverse trajectories approach (in negative time)
unstable stationary points.
We will conclude with a summary and brief discussion.

\section{Forward Projective Integration}

Projective integration was introduced in \cite{cwgygk}.
Its first-order form when used with the estimation of the chord slope
from a stochastic integrator is shown in \f{fig2}.

In practice we need to use a higher order method.
Here we will use an explicit multistep method similar to
the Adams-Bashforth method, which can be found in any standard
textbook \cite{gearbook}.
The third-order method is \eq Y(T+H)=
Y(T) +{{H}\over{12}}(23\dot{Y}(T) - 16\dot{Y}(T-H) + 5
\dot{Y}(T-2H)) \en where $H$ is the step size.
This formula
assumes that we have estimates of the derivatives $\dot{Y}$ that
are at least second-order accurate at three places: $T$, $T-H$,
and $T-2H$.
Higher order methods require correspondingly more
derivative estimates of correspondingly higher accuracy ($p$-th
order accurate estimates for an order $p+1$ integration method).
The formula also assumes that the spacing, $H$, between the
integration points is constant.
Neither of these assumptions is
strictly correct in the stochastic chord slope estimator we are
using.
Referring to \f{fig1} we see that if the microscopic
simulation starts at the point $T$, we will actually estimate the
slope of the chord between $T + n_1\delta t$ and $T + (n_1 + mn_3)
\delta t$.
This is a first-order accurate estimate of the
derivative at $T + (n_1 + mn_3/2)\delta t$, not at $T$.
Hence, we
need to modify the Adams-Bashforth method in two ways.

To describe the modifications, let us define $T_{n+1} = T_n + H$
where $H$ is the ``projective step size,'' that is, the distance
between successive groups of inner steps.
Let $T_n$ be the center of the chord computed via a least-squares fit.
Let $S_n$ be the computed slope of that chord.
Then we need to integrate from
$T_n$ to $T_n + H - (n_1+mn_3/2)\delta t$ to get an approximation
to the $Y$ value at the start of the next stochastic computation
as shown in \f{fig1}.
This is our first modification to the Adams-Bashforth method.
The first order method, which is Euler's
method, is straightforward: \eq Y(T_n + H_0) \approx Y(T_n) +
H_0S_n \en where $H_0 = H - (n_1+mn_3/2)\delta t$.
As long as
$S_n$ is a zeroth-order accurate approximation to $\dot{Y}(T_n)$,
this is a first-order method.
Second and higher-order methods must
reflect the fact that the final step is less than the spacing,
$H$, between the past points.
For example, the second-order
method is \eq Y(T_n + H_0) \approx Y(T_n) +
H_0(S_n+{{H_0}\over{2H}}(S_n - S_{n-1})). \en
If $S_n$ and
$S_{n-1}$ are first-order accurate approximations of the
derivative, this is a second-order method.
Similar modifications
apply to higher-order methods, and can be found via routine
algebra.

The second modification is due to the fact that we do not compute
approximations of the derivatives, but of the slopes of a
least-squares linear fit.
It is clear that these are first-order
approximations to the derivative at the center, so no further
adjustment is needed to these integration formulae through second
order.
However, the order of the difference between $S_n$ and
$\dot{Y}(T_n)$ is O($mn_3\delta t)$
(ignoring the stochastic noise).
The error thereby
introduced is of order O($H^2(mn_3\delta t/H))$.
In practice, whether or not an
additional correction is needed depends on the size of $mn_3\delta
t/H$.
If it
is small, the additional error is unimportant, but
if the projective step is not moderately large compared to the
length of microscopic integration used to estimate the chord
slope, additional corrections to the integration formula are
needed if the integration order is higher than two.

Figures \p{plot1} and \p{plot2} compare the results of projective
integration for a trajectory in the vicinity of the periodic
oscillations
observed at $\beta=20.8$, using $n_1\delta t=0.0175$, $mn_3\delta
t=0.005$, and $H_0=0.02$ ($H=0.04$) with an accurate integration using ODESSA.
The length of the ``settle time", $n_1\delta t$, was based on the
size of the fastest time constant corresponding to an eigenvalue around -6.
The projective integrator was based on the second-order
Adams-Bashforth methods.
The order, as well as the projective step size,
$H$, were chosen via the comparisons between different order and
step sizes shown in \t{T1}.
The values in this table were
computed as follows:  a point, $(\theta_A, \theta_B, \theta_C)$,
in the interior of the attractor in each of the  three
projected phase-plane
plots (as shown in \f{plot2} for the $\theta_A-\theta_B$
projection) was chosen.
Then, the difference
between the distances from this point to the ``true'' limit cycle
(as computed by ODESSA) and to the result of a stochastic
integration was computed at each integrated output point (or at a
sampling of them where they were densely packed).
A plot of one such set of errors is shown in \f{plot2a}.
The norm of this error
was computed over one limit cycle in each phase plane and the
average from the three 
projected
phase planes was used as the error estimate
in \t{T1}.

\begin{table} [t] \centerline{ \begin{tabular}{|c||c|c|c|c||} \hline
Projective Step size & Euler & AB 2nd & AB 3rd & AB 5th\\ \hline \hline
0.02 & 0.00025 & 0.00015 & 0.00014 & 0.00017\\
 0.04 & 0.00102 & 0.00086 & 0.00078 & unstable \\
0.06 & 0.00205 & 0.00221 & 0.00301 & unstable \\ \hline \hline
 \end{tabular} }
 \caption {Average radial deviation from the
attractor, for different time-steps and different projective integration
methods.}
 \label{T1}
\end{table}

\section{Coarse Limit Cycle Computation}

When we wish to compute the limit cycle from the output of a
coarse time stepper, the main challenges are the noisy nature of
the trajectories given by the coarse time-stepper and the fact
that the time stepper can only be iterated forward in time.

In order to compute a limit cycle (i.e., write a fixed point algorithm
that will converge on the limit cycle), we use the Poincar\'e map of
the trajectory.
The Poincar\'e map is defined by the successive
intersections of the time-stepper trajectory with a codimension-1
surface (for this 3D problem we choose conveniently a plane
$P(\vec{Y})$) transversal to the flow.
For the Poincar\'e map, a stable periodic trajectory of the original
flow becomes an attracting fixed point.

Assuming that $P$ is defined by $ \{\vec{Y} \in R^n: Y_{j}=C\}$ for
the $j-th$ variable, a
crossing of the Poincar\'e plane can be detected when the
quantity $(C-Y_j)$ changes sign \cite{Henon}.
One is only interested in crossings of the surface by the trajectory
{\it in the same direction}.
For noisy systems, care needs to be taken to avoid
the detection of false crossings.
Uncertainty in any state variable renders the computation of sign
changes on such variables unreliable.
However, in the presence of sufficient variance reduction, we
expect that the ensemble average of many stochastic realizations
to be much more reliable, and devise the
following procedure to detect a crossing of the Poincar\'e surface:

\begin{enumerate}
\item{Monitor the difference $|Y_j-C|$ along the trajectory.}
\item{If $|Y_j-C| < \delta$ at some time
($t_0$), gather a fixed number of points (M, until
$t=t_0+ M \Delta t$) along
the trajectory. With these points, approximate the trajectory of the
state variables ($k=1,2,...,n$) via a linear mapping:
$Y_k=a_kt+b_k$.}
\item{Assisted by this local model, confirm that a crossing has occurred.
That is, $Y_j=C$ for some  $t\in [t_0,t_0+ M \Delta t]$.
If this condition is not
met, return to step 1.}
\end{enumerate}

\noindent
Note that this algorithm also gives information about the direction
of the crossing (the sign of the slope of the local model for $x_i$);
thus, the relevant crossings at the Poincar\'e surface (the ones
having the same sign of $a_i$) are obtained.

In the neighborhood of a simple stable periodic trajectory,
the dynamics on the Poincar\'e section will appear as a sequence
of points approaching the fixed point.
Implementing a Newton-Raphson-type contraction
mapping on the coarse return map,
the convergence to the fixed
point of the sequence (i.e., to the periodic trajectory of the
flow) can be accelerated.
This contraction map is defined as
follows. Let {\bf x} be the vector of components of $Y$ excluding
$Y_i$ used to define the Poincar\'e section.

$${\bf x}^{k+1}= {\bf x}^{k} =- J({\bf x}^k)^{-1}({\bf x}^k-{\bf x}^{k-1})$$

\noindent
where $J$ is the Jacobian of the linearization of the Poincar\'e map
(corresponding to the monodromy matrix for the limit cycle).

The Jacobian can be estimated by applying perturbations around
the current Poincar\'e crossing.
A ``centered difference" ensemble of perturbations are used to estimate
numerical derivatives, which are naturally sensitive to noise.
For problems with many coarse variables, Newton-Krylov type methods
based on coarse timesteppers are being explored (such as the Recursive
Projection Method of Shroff and Keller, or
Newton-Picard methods \cite{rpm,lust}); the sensitivity of these
matrix-free methods to noise is an important focus of current
research.

\subsection{Results}

The kMC simulations were performed as described in Section 2 using
$N=(1000)^2$ adsorption sites and computing the average over 100
realizations for $\beta=20.24$
\footnote{$\beta = 20.24$ was used for the fixed-point iteration
because the differential equation solution converges very slowly to
the limit cycle, thus making the problem more challenging.}.
The deterministic (mean-field approximation) system
exhibits, at these conditions, a single, attracting
periodic trajectory.
Observing on a Poincar\'e map defined by the plane $\theta_A=0.33$
and considering only crossings with negative slope, the fixed point of
the Poincar\'e section is found at $(\theta_B,\theta_c)
=(0.027947,0.61173)$.
It is indicated with a filled square in
\f{plot6}.
In addition to the ubiquitous Floquet multiplier at 1,
the leading multiplier is 0.72147 while the second
multiplier is $2.92\times10^{-8}$. The return time is 183.89 time
units. For the selected value of $\beta$, trajectories converge
very slowly on the limit cycle.

\subsubsection{{\bf Converging to the Limit cycle}}

Note the drastic time-scale separation implied by the
orders of magnitude difference of the Floquet multipliers
of the deterministic, mean field limit cycle.
This suggests that perturbations along
{\it one} direction of the two-dimensional Poincar\'e section
decay very quickly.
In order to avoid large numerical sensitivity
during the Jacobian estimation, we exploit this separation of
time scales by resorting to an effectively one-dimensional
approach.
In the time interval of even a single return time, we can
consider that $\theta_C$ at the Poincar\'e
crossing becomes slaved to $\theta_B$.
This fast slaving is ``embodied" in the eigenvector of the
fast (strongly stable) Floquet multiplier (we can ``read" it
in the corresponding eigenvector of the linearization around
the fixed point for the deterministic
-- mean-field -- system).
The resulting line equation (see
Fig.~\ref{plot6} all points of the NR iteration lie on it) is:

$$\theta_C=(1.0-2.63149316\theta_B)/1.51747895$$

Performing the contraction mapping computations in the remaining
one dimension converges to the coarse fixed point
with an an error $O(10^{-4})$, commensurate with the
expected uncertainty level from the kMC simulation.
Our convergence criterion (maximum deviation)
was set to 0.0001 on the coarse iteration variable $\theta_B$.
Our contraction mapping converges after 4
iterations (point 0 on Fig.~\ref{plot6} is the initial point).
The
derivatives of the map were estimated using three perturbations
around the current point and fitting a least-squares line (3
points); the perturbations were 0.0005, 0.0, and -0.0005 on
$\theta_B$.
The estimated coarse leading multiplier was 0.7274 and the
coarse return time 188.19 time units; the location of the
coarse fixed point is estimated to be at $(0.0282,0.6101)$,
all in good agreement with the mean-field approximation.

\section{Reverse Projective Integration}

Reverse projective integration, described in \cite{backward}, is a
version of projective integration in which the inner integrator
proceeds in the forward direction while the projective step is taken in
the reverse direction.
It was used in \cite{backward} because of
its unusual stability properties that enabled us to find
certain classes of stationary points.
Here we use it to integrate backward in time when the
microscopic system can only be evolved
forward in time, either because of its nature or because
it is defined by a legacy code which cannot easily be modified.

Reverse integration is illustrated in
\f{newfig3} with a linear projective step.
In \cite{backward} it was shown that reverse projective
integration damps components corresponding to large negative
eigenvalues - components that would be unstable in normal reverse
integration were the latter possible.
The reverse projective step causes the
method to also be damping for components corresponding to small
positive eigenvalues since these decay in reverse time.
Thus
reverse projective integration
allows us to approach stationary points that are stable or unstable
(including {\it saddle} points),
provided that the local eigenvalues in the negative
half plane have large negative real parts while any in the
positive half plane have small real parts.

When the
microscopic model is stochastic, it is usually inherently
uni-directional in time.
For example, a random walk will give rise to effective diffusion,
even if the sign of all movements (and thus, time) was reversed.
This has been exploited in \cite{Hummer} to guide molecular
dynamics simulations backwards on a free energy surface, and
in the location of transition states.
Here, we
will combine a stochastic microscopic model with a reverse
projective integrator to approach, backward in time,
a stationary saddle point of a KMC model.

Note that many types of fixed points can be located performing
contraction mappings ``wrapped around" the coarse timestepper
(e.g. the Shroff and Keller RPM \cite{rpm}).
That procedure, however, requires a good initial condition.
Reverse integration gives approximations to backward in time
trajectories that will be attracted (in reverse time) to
these forwardly unstable points.

To illustrate this method we first use a further simplification of
\r{alexieq} to a single ODE that has an unstable stationary point
(its sole eigenvalue is positive).
It is given in \cite{alexi} as
\eqn  {{d\theta}\over{dt}}=\alpha (1-\theta) -\gamma\theta-k_r
(1-\theta)^2\theta \enn{alexieq1} where now, $\alpha=1.0$, and
$\gamma=0.01$.
In this case, \r{alexieq1} can be integrated in
reverse time to find the unstable stationary point.
Reverse projective
integration can also be used to find that stationary point.
Fig.~\ref{plot3} shows the comparison between the reverse
deterministic and reverse projective-stochastic integrations for
a particular initial condition.
The parameter $k_r$ was set to 5. The
trajectories (starting at time=0.0) approach the unstable fixed
point at $\theta = 0.7357$ whose eigenvalue is about 0.58.
The deterministic trajectory was obtained from ODESSA \cite{odessa} with
a reporting horizon of -0.02 units and a tolerance of $1\times10^{-9}$.
The reverse projective method used
the stochastic code forward for 0.02 units of time and then
the Adams-Bashforth second-order formula, as described above, with
a time step of -0.12 time units.
As before, we used for this
example $10^6$ adsorption sites and the results are averaged over
100 simultaneous realizations.

Figure ~\ref{plot4} shows similar integrations
for the ``full" version, \r{alexieq}. The parameter $\beta$ is set
to 20.8.
The reverse trajectories
approach a saddle stationary point in reverse time.
The eigenvalues of the
linearization of the flow along the deterministic trajectory are
presented in \f{plot4eigen}(a) and (b).
In this case, the deterministic
trajectory was also calculated using reverse projective
integration with ODESSA as the inner integrator in the forward
direction.
(This was necessary because of the saddle nature of
the stationary point, meaning that direct reverse integration would be
unstable and explode backward in time.)
ODESSA was used to integrate for one unit of time in
the forward direction and then a projective Adams-Bashforth step was
used in
the reverse direction for two units of time.
We confirmed, by starting at the ``end" of the reverse trajectory,
and using ODESSA normally forward in time, that the reverse projective
trajectory was indeed traced by the code in forward time \f{plotconfirm}.
The projective
stochastic solution was calculated with the same steps, but the
forward evolution was carried out with the stochastic code.
Significant numerical noise arises along the stochastic trajectory.
This noise can be partially explained by
studying the linearization of the vectorfield (computed or estimated)
along the trajectory
and in particular the eigenvalues closer to zero.
A small perturbation along the
trajectory is amplified by the factors plotted in \f{plot4eigen}(c)
(these
are $e^{\lambda \delta t}$ for the largest $\lambda$, where $\delta t$
is the total reverse integration time step).
Even when such perturbations (that
are a natural part of the stochastic simulation results) are not
amplified, they decay very slowly.
The choice of time-step is critical, as we need to use a relatively long
forward
``inner" integration to damp the stiff eigencomponents which would
otherwise blow up the backward trajectory.
Fig.~\ref{plot4}(b) and (c) shows that the deterministic and the
stochastic
reverse projective trajectories are similar when plotted in phase-space.

Figure~\ref{plot4b}
shows the same results when the reverse trajectory is initiated closer
to the
saddle point, beyond the region where perturbations are amplified
(at the point of the deterministic trajectory of Fig.~\ref{plot4}
where time=-180 s).
The agreement is better, although still not noise-free.
The important point is that the reverse trajectory
asymptotically tends to the saddle point.

\section{Conclusion}

We have demonstrated the use of projective integration techniques for
simulating the coarse dynamic behavior of models described by microscopic
models.
In this paper, the ``inner" microscopic model was a kinetic Monte Carlo
simulation of surface reaction schemes based on the Stochastic Simulation
Algorithm of Gillespie.
It is worth noting more recent work by Gillespie on the so-called
``tau-leaping" method, which can be used under some circumstances to
accelerate stochastic simulation algorithms (SSAs)  \cite{tau_leaping1}.
Our coarse projective integration schemes are computational ``wrappers" around
``inner evolution" schemes; as such, they can be wrapped around
a ``tau-leaping" inner SSA.
Implicit versions of coarse projective integrators have been introduced in
\cite{cwgygk}; studying the analogies and differences between these
methods and the ``implicit tau-leaping" schemes of Rathinam et al.
\cite{tau_leaping2} is a
subject of current research.

Beyond the illustration of coarse projective integration in a kinetic
Monte Carlo context,  the focus of this
paper was in the use of such schemes to perform tasks beyond direct simulation.
Reverse coarse projective integration was demonstrated; its ability,
under certain
circumstances, to approach unstable, saddle-like objects in coarse phase space
may prove helpful in the location of transition states
in computational chemistry
(see the coarse MD example in \cite{Hummer}).
We also illustrated the use of projective integration in the solution
of coarse boundary value problems, in particular the location and computer-assisted
stability analysis of coarse limit cycles for the expected behavior of the
microscopic simulator.

Remarkably, these algorithms have counterparts in the case of continuum numerical
analysis also; these counterparts are particularly meaningful in the context of
accelerating legacy simulators for continuum problems (see for example the stability
analysis in \cite{cwgygk} and \cite{backward}).
They also constitute the inspiration for further algorithms, like telescopic projective
integrators \cite{cwgygk} for problems with several gaps in their eigenvalue spectrum,
as well as coarse Langevin-based acceleration techniques for problems whose dynamics
are governed by rare events.
The extraction of macroscopic dynamics from microscopic/stochastic simulators
constitutes the current frontier in multiscale/complex system computation
\cite{Stuart}.
Equation-free techniques, based on coarse time-stepping, aim at {\it solving} these
macroscopic equations without ever deriving them in closed form; as such, techniques
like the coarse projective integration illustrated here constitute a bridge between
traditional continuum numerical analysis and detailed physical modeling of complex systems.

{\bf Acknowledgements} This work was partially supported by AFOSR
(Dynamics and Control),
an NSF-ITR grant (IGK, CWG) and Fulbright- Garc\'{\i}a Robles
and CONACYT Fellowships (RRM).
Discussions with Profs. Li Ju,
P. G. Kevrekidis, L. Petzold and Dr. G. Hummer are gratefully acknowledged.

\newpage   

\begin{figure}[ht]
\centerline{\psfig{file=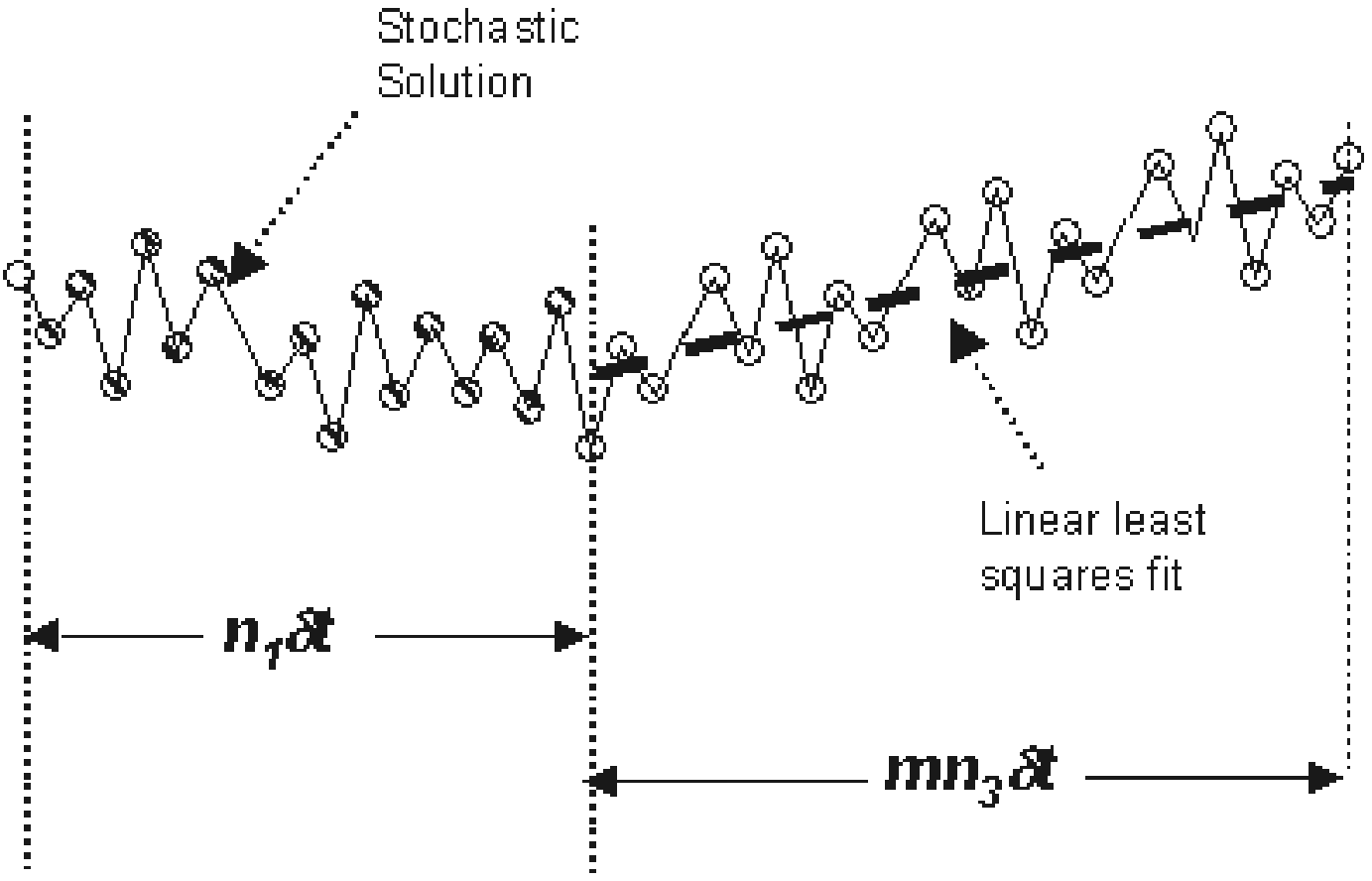,height=2in,width=4.0in,angle=0}}
\caption{Schematic of the coarse chord slope computation.} \label{fig1}
\end{figure}

\begin{figure}[ht]
\centerline{\psfig{file=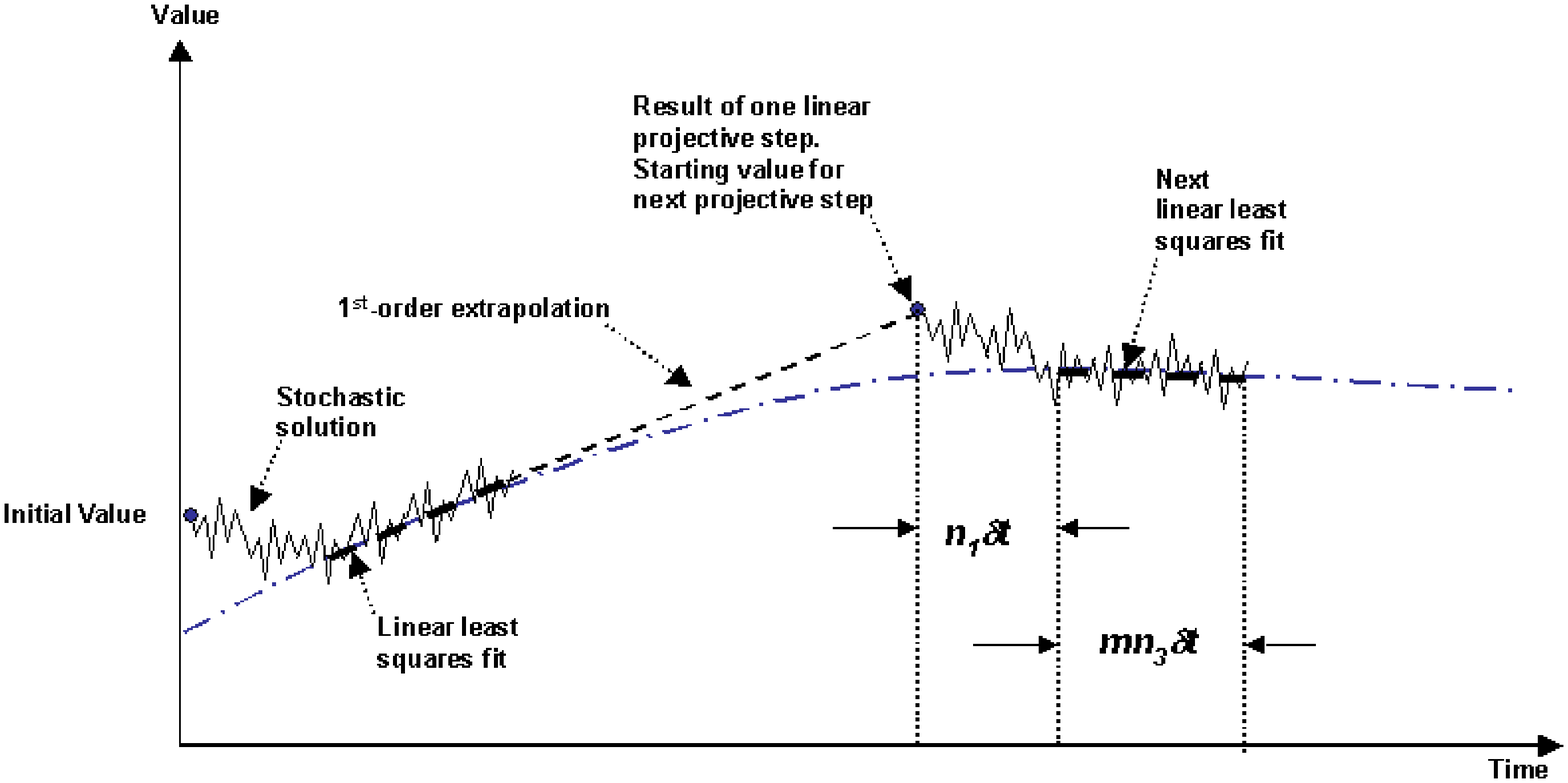,height=2.5in,width=4.0in,angle=0}}
\caption{Schematic of first-order coarse projective integration.}
\label{fig2}
\end{figure}

\begin{figure}[ht]
\centerline{\psfig{file=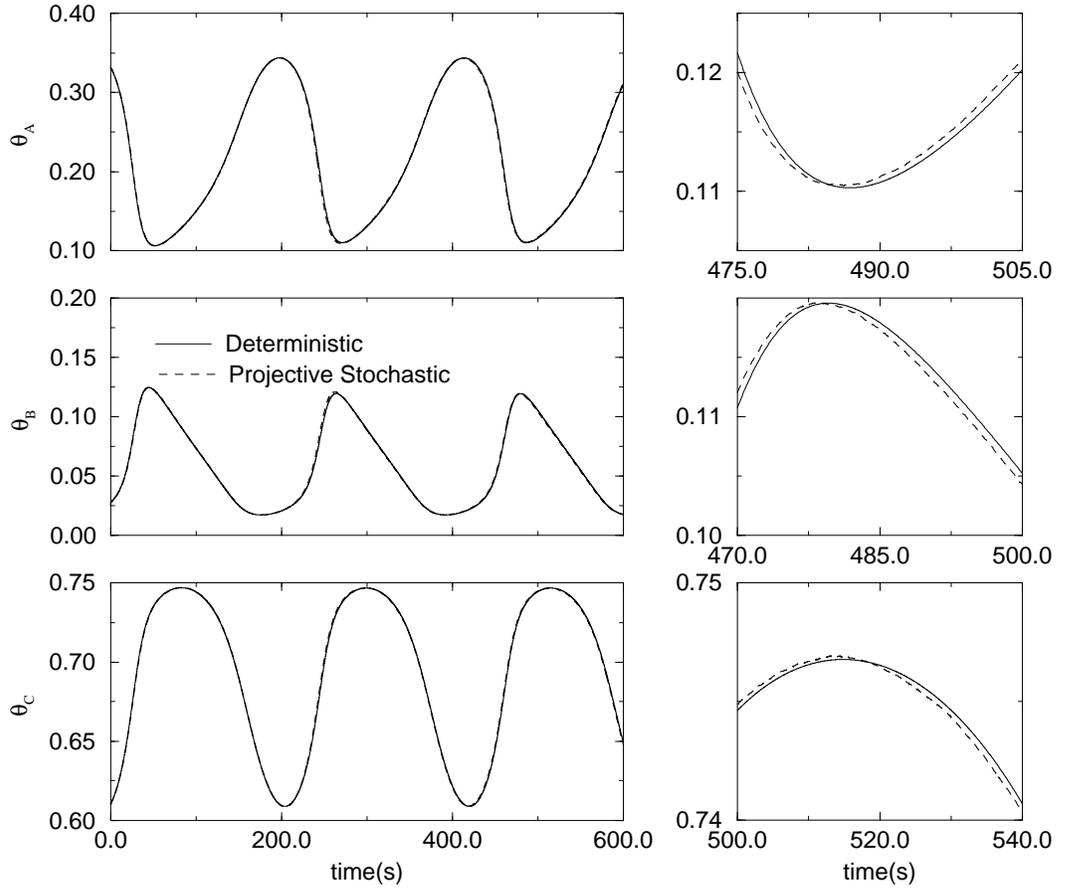,width=5.0in,angle=270}}
\caption{An illustration of forward coarse projective integration.
Periodic trajectory of the model system
computed using the mean-field equations (solid line), and the kMC simulator
(dashed line).
A second-order Adams-Bashforth integrator was used to compute the kMC
trajectory with $n_1\delta t=0.0175$, $mn_3\delta
t=0.005$, and $H_0=0.02$ ($H=0.04$) for the projective algorithm.
For this example $\beta=20.8$, other parameters of the model were set
as indicated in the text.}
\label{plot1}
\end{figure}

\begin{figure}[ht]
\centerline{\psfig{file=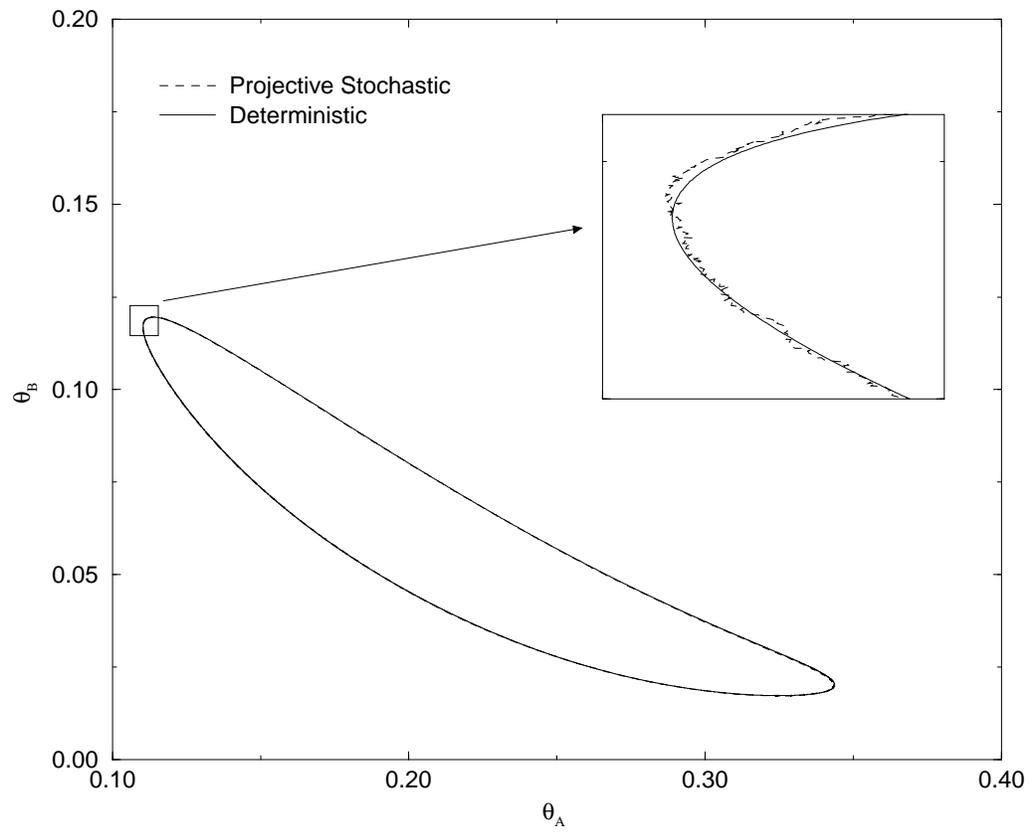,width=5.0in,angle=270}}
\caption{Attractor comparison in the $\theta_A-\theta_B$
projection for the results presented in Fig. 3.}
\label{plot2}
\end{figure}

\begin{figure}[ht]
\vspace{+0.5in}
\centerline{\psfig{file=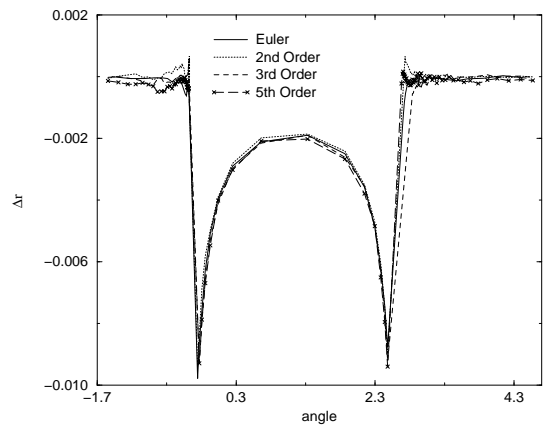,width=3.0in,angle=270}}
\vspace{-.5in} \caption{Local errors (angle parametrization,
around the point (0.22,0.07,0.7))
for the $\theta_A-\theta_B$
phase plane projection of the attractor. Results of several integrator orders
using the example illustrated in Fig. 3 are presented.} \label{plot2a}
\end{figure}

\begin{figure}[p]
\centerline{\psfig{file=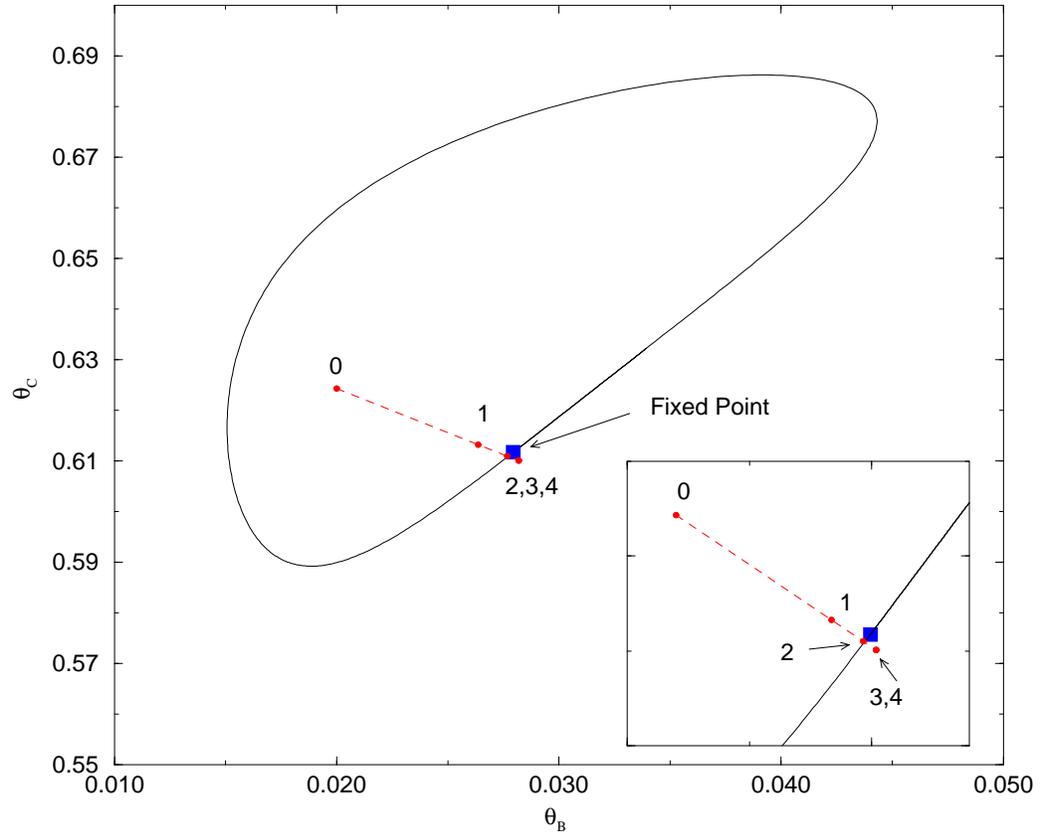,width=5in,angle=270}}
\caption{Fixed point iteration and convergence to the
coarse limit cycle
with $\beta = 20.24$.
The initial point is marked 0, remaining points
are given by a coarse Newton-type iteration formula.}
\label{plot6}
\end{figure}

\begin{figure}[ht]
\centerline{\psfig{file=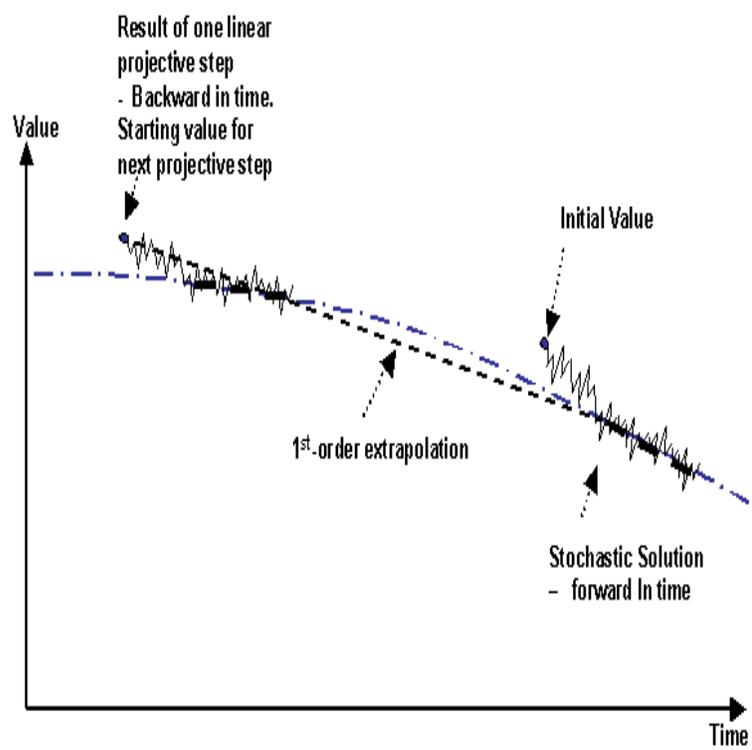,height=4in,width=4.0in,angle=0}}
\caption{Schematic of reverse coarse projective integration.}
\label{newfig3}
\end{figure}

\begin{figure}[t]
\centerline{\psfig{file=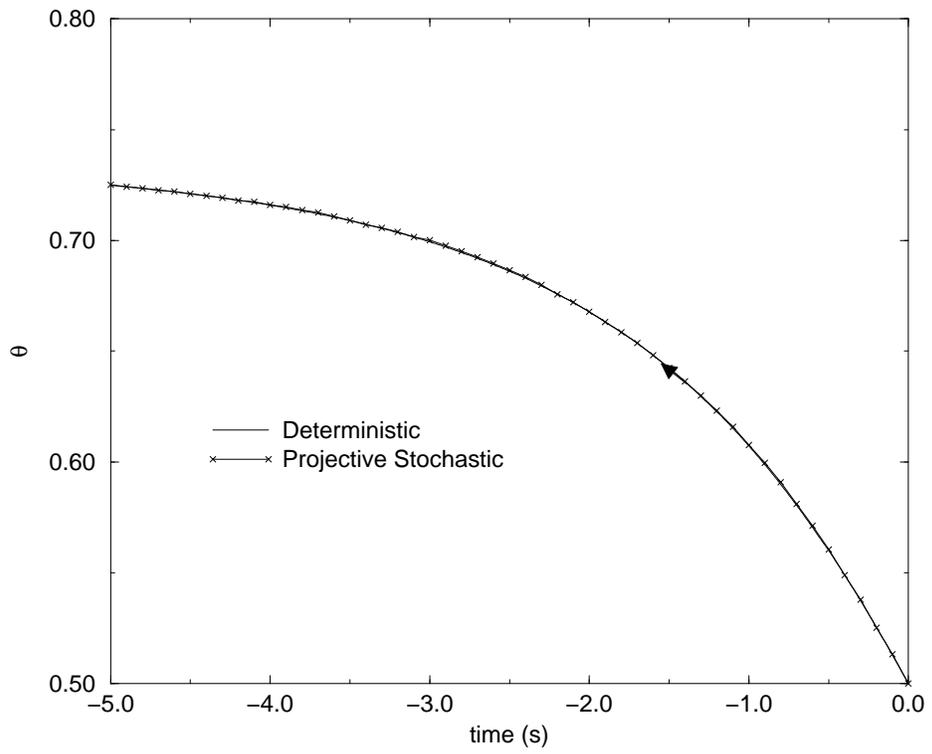,width=5.0in,angle=270}}
\caption{Backward projective integration for the simplified model.
The trajectory approaches
an unstable fixed point.}
\label{plot3}
\end{figure}

\begin{figure}[ht]
\centerline{\psfig{file=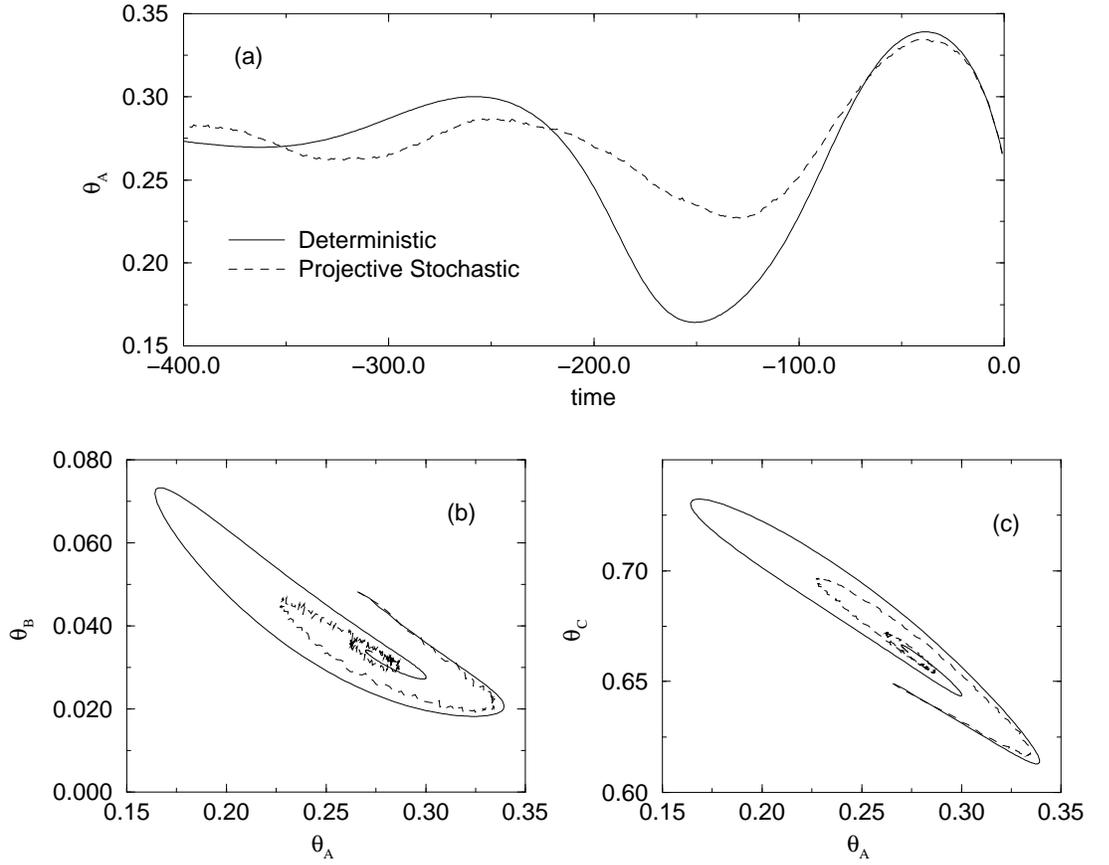,width=5in,angle=270}}
\caption{Backward integration for the full model
with $\beta  = 20.8$.
(a) Time-series of
$\theta_A$ evolution, (b) and (c) Projections of the trajectory
in phase-space. The parameters of the projective integrator, and
other parameters
of the model, are described in the text. }
\label{plot4}
\end{figure}

\begin{figure}[ht]
\centerline{\psfig{file=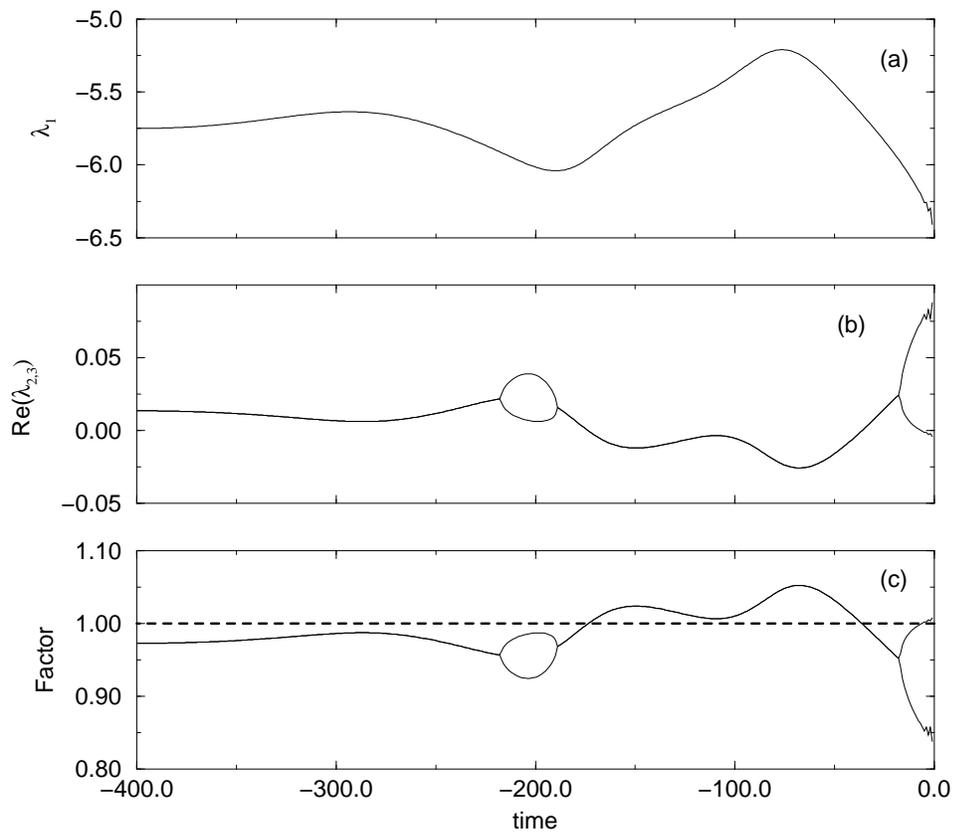,width=5in,angle=270}}
\caption{Backward integration for the full model. (a) and (b)
eigenvalues along
the trajectory. (c) amplification factor for perturbations
at various locations along the trajectory. }
\label{plot4eigen}
\end{figure}

\begin{figure}[ht]
\centerline{\psfig{file=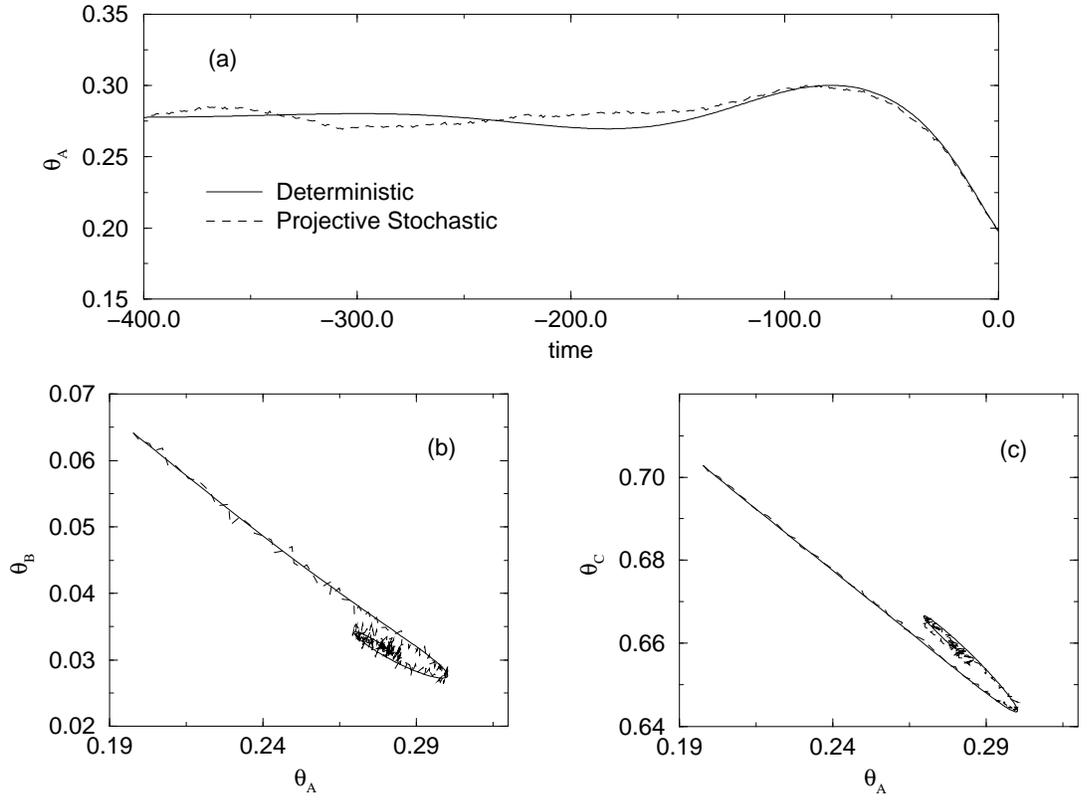,width=5in,angle=270}}
\caption{Backward integration for the full model. (a) Time-series of
$\theta_A$ evolution, (b) and (c) Projections of the trajectory
in phase-space.}
\label{plot4b}
\end{figure}

\begin{figure}[ht]
\centerline{\psfig{file=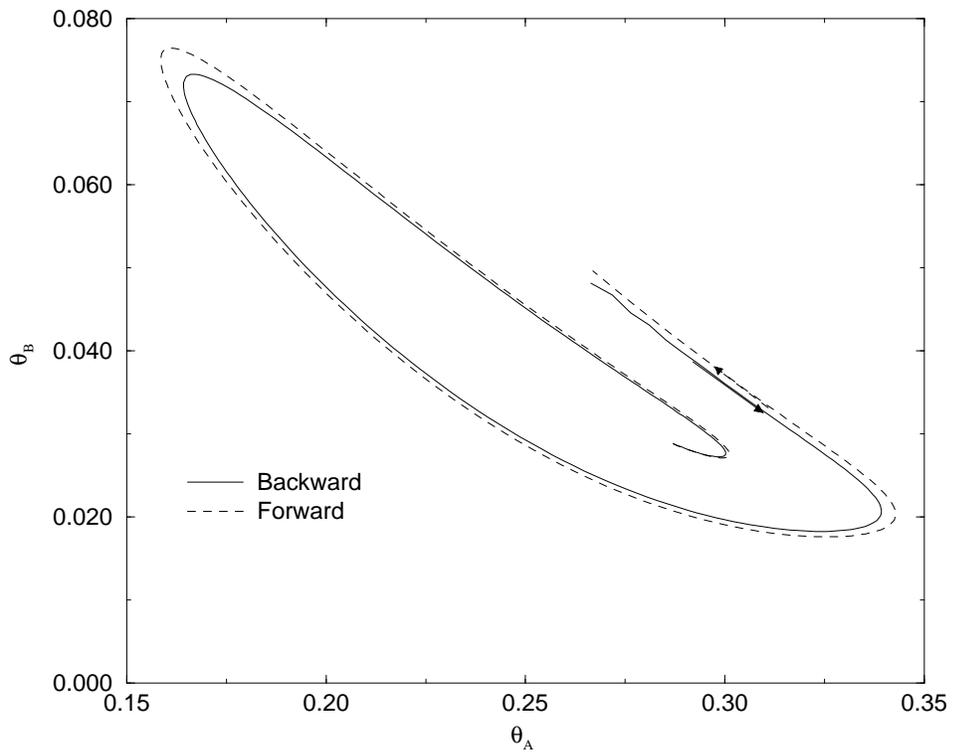,width=5in,angle=270}}
\caption{Comparison of backward and forward integration for
the example in Fig. 9.
The forward integration trajectory (dashed line) was
obtained by using, as a starting point, the final point of the backward
integration.}
\label{plotconfirm}
\end{figure}

\end{document}